\documentclass[modern]{aastex631}

\begin{document}

\title{HD 110067 is a wide hierarchical triple system}

\correspondingauthor{Kevin Apps}
\email{kevinapps73@gmail.com}

\author{Kevin Apps}
\affiliation{Independent scholar, UK}

\author[0000-0002-4671-2957]{Rafael Luque}
\affiliation{Department of Astronomy \& Astrophysics, University of Chicago, Chicago, IL 60637, USA}



\begin{abstract}

We report that HD 110067, the recently announced host star of a resonant sextuplet of transiting sub-Neptunes, is not a single star as claimed in the discovery paper, but a wide hierarchical triple. The K0\,V planet hosting star ($V = 8.4\,\mathrm{mag}$, $d = 32\,\mathrm{pc}$) has a companion at a wide projected separation of 13400\,au. This companion, namely HD~110106, is a slightly fainter ($V=8.8\,\mathrm{mag}$) K3\,V type 8-year period double-lined spectroscopic binary. The secondary in this spectroscopic binary is contributing a significant amount of flux and has a measured high mass ratio.

\end{abstract}



\section{The bright companion to HD 110067} \label{sec:companion}

\citet{Luque2023Natur.623..932L} have recently announced the discovery of six sub-Neptunes in a chain of first-order mean motion resonances transiting the nearby star HD 110067. The authors used high-resolution imaging observations from PHARO at Palomar Observatory and Alopeke at Gemini North to exclude the presence of close-in ($\lesssim 1$\,arcsec) companions. Although HD~110067 does not have any close companion, the authors incorrectly assumed that the star is single.

The Washington Double Star Catalogue \citep{WDS} lists HD 110067 as a wide binary (BGH 40 AB) with 17 observations spanning from 1893 to 2015. The companion is HD 110106, classified as K3\,V by \citet{YossGriffin97} and with a magnitude of $V=8.8\,\mathrm{mag}$. The separation of this companion at epoch 2016.0 according to \textit{Gaia} DR3 \citep{GaiaDR3} is $415.701\pm0.001\,\mathrm{arcsec}$ at $\mathrm{PA} = 148.1\,\mathrm{deg}$. Using the \textit{Gaia} DR3 parallax of the primary ($\pi = 31.037\pm0.022\,\mathrm{mas}$), this corresponds to a projected separation of $13394\pm10$\,au.  For reference, the apastron of our nearest neighbour Proxima Centauri from Alpha Centauri AB is estimated to be a similar $13300^{+300}_{-100}$\,au \citep{Kervella2017}. Table~\ref{tab:binary} shows the main astrometric properties of HD 110067 and HD 110106.

\begin{table}[t]
\caption{Astrometric properties of the wide binary pair BGH 40 AB.}\label{tab:binary}
\centering
\begin{tabular}{lccccr}
\hline
\hline
\noalign{\smallskip}
Star & $\pi$ (mas) & $\mu_{\alpha}\cos\delta$ ($\mathrm{mas/yr}$) & $\mu_{\delta}$ ($\mathrm{mas/yr}$) & RV ($\mathrm{km/s}$) & References \\
\noalign{\smallskip}
\hline
\noalign{\smallskip}
HD 110067 & $31.04\pm0.02$ & $-81.70\pm0.03$ & $-104.53\pm0.02$ & $-8.56\pm0.13$    & \textit{Gaia} DR3 \\
\noalign{\smallskip}
HD 110106 & $28.13\pm0.31$ & $-85.07\pm0.40$ & $-102.34\pm0.38$ & $\cdots$          & \textit{Gaia} DR3 \\
~~~~~~~~~ & $\cdots$       & $-82.20\pm1.00$ & $-103.90\pm1.00$ & $\cdots$          & Tycho-2 \\
\noalign{\smallskip}
\hline
\end{tabular}
\end{table}

\section{The binary nature of HD 110106} \label{sec:companion}

While the planet host star, HD 110067, has a low Gaia Renormalised Unit Weight Error (RUWE) of 0.943, the RUWE of its companion, HD 110106 is 14.878. Values greater than 1.4 indicate excessive astrometric noise, typically caused by the presence of an unseen companion \citep{StassunTorres21}. The poor fit on the astrometric solution is reflected in Table~\ref{tab:binary}, where the formal error bars on the parallax and proper motion from \textit{Gaia} DR3 are over 10 times higher than its otherwise similar companion, with the parallax being 9.5$\sigma$ different. The values from the Tycho-2 catalogue \citep{Tycho2}, however, agree within 1$\sigma$ on the proper motions with the primary. This is due to the longer baseline of the Tycho 2 catalog (spanning several decades) compared to the less than 3-year astrometric baseline of \textit{Gaia}. The binary orbit being less than three times the length of the \textit{Gaia} astrometric baseline is the main reason behind the poor astrometric fit of the secondary. Future \textit{Gaia} releases including epoch astrometry will enable a joint spectroscopic and astrometric fit to refine the orbital properties of the system.

HD 110106 is a known spectroscopic binary identified by \citet{Halbwachs12} using 22 CORAVEL observations spanning 3884 days. The orbit has the following parameters: $P = 2899.2\pm20.5$\,d, $T_0 = 2446789.4\pm33.7$\,d, $e = 0.260\pm0.017$, $\omega = 129.90\pm3.84$\,deg, $K = 6.821\pm0.152$\,km/s, and $\gamma = -8.555\pm0.087$\,km/s. The RMS of the fit is 0.336\,km/s. Later observations using SOPHIE \citep{Halbwachs14} enabled the detection of the spectra lines of the two components. The spectroscopic RV amplitude ratio ($K_1/K_2 = 0.843\pm0.056$) suggest a high mass ratio between components.  This is supported by the relative positions of the primary and secondary compared to the mean local main sequence in color-magnitude diagrams\footnote{These offsets have been determined by the lead author's own 800 MB database of nearby stars, assuming the parallax of the system is equal to the value of HD 110067. These color-magnitude diagrams are available on request from the author.}. The offsets in \textit{Gaia} $B_G-R_G$ vs. $M_G$ is -0.15\,mag for the primary and +0.44\,mag for the secondary. The offsets are similar in hybrid \textit{Gaia} and \textit{2MASS}, $G-K$ vs. $M_K$ offset of -0.17\,mag for the primary and +0.51\,mag for the secondary. Combining both lines of evidence (spectroscopic amplitude ratio and the overluminosity of the secondary), it seems likely that HD 110106 is a nearly equal binary.



\bibliography{sample631}{}
\bibliographystyle{aasjournal}



\end{document}